\documentclass[twocolumn,showpacs,prd,floatfix,axodraw]{revtex4}
\usepackage{mathrsfs}
\usepackage{graphicx,,booktabs,bm}
\usepackage{overpic}
\usepackage{color}
\usepackage{amssymb}

\usepackage{mathrsfs,bm,amsmath,amssymb}
\usepackage{longtable,lscape}
\usepackage{txfonts}
\usepackage{amssymb}
\usepackage{indentfirst}
\usepackage{graphicx,,booktabs}
\usepackage{multirow}
\usepackage{color}
\usepackage{amssymb}

\definecolor{cover}{rgb}{0.77,0.87,0.88}
\definecolor{blueone}{rgb}{0.1,0.1,.7}
\definecolor{citec}{rgb}{0.14,0.47,0.09}
\definecolor{two}{rgb}{0.0,0.5,0.}
\definecolor{three}{rgb}{.5,.1,0.15}
\usepackage[bookmarks=true,bookmarksopen=false,plainpages=false,breaklinks=true,
   bookmarksnumbered=true,hypertexnames=false,
   filecolor=blue,urlcolor=three,menucolor=three,
   linkcolor=three,citecolor=blueone, colorlinks,
   anchorcolor=blue,runcolor=pink,frenchlinks=red
   pdfstartview=FitH,pdftitle=title,%
   pdfauthor=author]{hyperref}

\begin{document}
\title{{Radiative decay of the $\Xi_c(2923)$ in a hadronic molecule picture}}

\author{ Feng Yang}
\affiliation{ School of Physical Science and Technology, Southwest Jiaotong University, Chengdu 610031,China}

\author{Hong Qiang Zhu}
\email{20132013@cqnu.edu.cn}
\affiliation{College of Physics and Electronic Engineering, Chongqing Normal University, Chongqing 401331,China}

\author{Yin Huang} 
\affiliation{Asia Pacific Center for Theoretical Physics, Pohang University of Science and Technology, Pohang 37673,
Gyeongsangbuk-do, South Korea}
\affiliation{School of Physical Science and Technology, Southwest Jiaotong University, Chengdu 610031,China}

\date{\today}
\begin{abstract}
In the present work,  we study the radiative decay of newly observed $\Xi_c(2923)^0$ based on the successful explanation that $\Xi_c(2923)^0$
is an $S$-wave $D\Lambda-D\Sigma$ molecular state in our previous study~\cite{Zhu:2020jke}.  The radiative decay width of $D\Lambda-D\Sigma$
molecular state into $\Xi_c^0\gamma$ final state through hadronic loops are evaluated using effective Lagrangians.  We find that decay widths
$\Xi_c(2923)^0\to\Xi_c^0\gamma$ and $\Xi_c(2923)^0\to\Xi_c^{'0}\gamma$ is evaluated to be approximately 1.23-11.66 KeV and 0.30-3.71 KeV, respectively.
These are different from the results~\cite{Wang:2020gkn,Bijker:2020tns} that obtained by assuming $\Xi_c(2923)^0$ may be conventional
charmed baryon.  If measurements are in future experimental, these differences will be very useful to help us to test various interpretations of $\Xi_c(2923)^0$.
\end{abstract}


\maketitle
\section{INTRODUCTION}
At present, there are thirty baryons with only one charm quark listed in the review of PDG~\cite{Zyla:2020zbs}.
Understanding their internal structure is one of the most meaningful topics in particle and nuclear physics.
Among them, three newly observed neutral resonance $\Xi_c^{*0}$ named $\Xi_c(2923)^0$, $\Xi_c(2939)^0$, and
$\Xi(2965)^0$ in the $K^{-}\Lambda_c^{+}$ mass spectra by the LHCb Collaboration~\cite{LHCb:2020iby} aroused
widespread discussion. Their masses and widths are
\begin{align}
\Xi_c(2923)^0: ~M&=2923.04\pm{}0.25\pm 0.20\pm 0.14~~~{\rm MeV}\nonumber\\
\Gamma&=7.1\pm{}0.8\pm 1.8~~~~~ {\rm MeV},\nonumber\\
\Xi_c(2938)^0: ~M&=2938.55\pm{}0.21\pm 0.17\pm 0.14~~~{\rm MeV}\nonumber\\
\Gamma&=10.2\pm{}0.8\pm 1.1~~~~~ {\rm MeV},\nonumber\\
\Xi_c(2964)^0: ~M&=2964.88\pm{}0.26\pm 0.14\pm 0.14~~~{\rm MeV}\nonumber\\
\Gamma&=14.1\pm{}0.9\pm 1.3~~~~~ {\rm MeV},\nonumber
\end{align}
respectively.

Due to the uncertainty of spin-parity, many disputes about their internal structure have been proposed.  In the chiral quark model,
based on the analysis of the two-body Okubo-Zweig-Iizuka (OZI) allowed strong decays, the states $\Xi_c(2923)^0$, $\Xi_c(2939)^0$,
and $\Xi(2965)^0$ can be considered as $1P$ $\Xi_c^{'}$ state with spin-parity $J^P=3/2^{-}$ or $J^P=5/2^{-}$~\cite{Wang:2020gkn}.
In this paper the authors suggested to search for them in the $\Xi_c(2923)^0/\Xi_c(2939)^0\to\Xi_c^{'}\pi$ reaction and the
$\Xi(2965)^0\to\Lambda_c{}K/\Xi_c\pi$ reaction that can well test the $dsc$ nature of the $\Xi_c^{*0}$.  By employing the $^3P_0$ approach
in Ref.~\cite{Lu:2020ivo}, the two-body strong decays of the $\Xi_c(2923)^0$, $\Xi_c(2939)^0$, and $\Xi(2965)^0$ is calculated.
The results indicate that the $\Xi_c(2923)^0$ and $\Xi_c(2939)^0$ can also be $1P$ $\Xi_c^{'}$ state, while the $\Xi(2965)^0$ is
suggested to be $2S$ $\Xi_c^{'}$ state.  The QCD sum rule suggests that the states $\Xi_c(2923)^0$, $\Xi_c(2939)^0$, and $\Xi(2965)^0$
are most likely to be considered as the $P$-wave $\Xi_c^{'}$ baryons with the spin-parity
$J^P=1/2^{-}$ or $J^P=3/2^{-}$~\cite{Yang:2020zjl}.    These are different from the results by Agaev et al.~\cite{Agaev:2020fut}
that the $\Xi_c(2923)^0$ and $\Xi_c(2939)^0$ can be considered as $1P$ excitations of the spin-1/2 flavor-sextet and spin-3/2
baryons, respectively, while the $\Xi(2965)^0$ may be the exciting 2S state of either spin-1/2 flavor-sextet or antitriplet baryon.
In addition, the resonance states of the five-quark configuration are possible candidates of these new states with negative parity~\cite{Hu:2020zwc}.
However, a completely different conclusion was drawn from Ref.~\cite{Zhu:2020jke} that the $\Xi_c(2923)^0$ can be understood
as a $S-$ wave $D\Lambda-D\Sigma$ bound state.   Indeed, a $D\Lambda$ or $D\Sigma$ abound state with a mass
about 2930 MeV that can be associated to the $\Xi_c(2923)^0$ is supported by Refs.~\cite{Jimenez-Tejero:2009cyn,Yu:2018yxl,Nieves:2019jhp}.
The lattice QCD calculation was also performed and tried to determine their quantum numbers~\cite{Bahtiyar:2020uuj}.

The successful explanation of $cds$ state~\cite{Wang:2020gkn,Lu:2020ivo,Yang:2020zjl,Agaev:2020fut,Hu:2020zwc} or
molecular structure~\cite{Zhu:2020jke,Jimenez-Tejero:2009cyn,Yu:2018yxl,Nieves:2019jhp} for newly observed $\Xi_c^0$ states
naturally leads us to ask what is their real structure?   Moreover, there still exist many incomprehensible conclusions, such as the
studies of conventional charm baryon cannot agree on their spin-parity~\cite{Wang:2020gkn,Lu:2020ivo,Yang:2020zjl,Agaev:2020fut,Hu:2020zwc}.
More important is that the two-body allowed strong decay widths from different models are consistent with each other within errors.
Hence, based on analysis of the two-body allowed strong decays, newly observed $\Xi_c^0$ states are not only considered as
conventional three quark state~\cite{Wang:2020gkn,Lu:2020ivo,Yang:2020zjl,Agaev:2020fut,Hu:2020zwc}, but also as $S$-wave molecular
configuration~\cite{Zhu:2020jke}.  Precise information on the radiative decay mechanism
of these $\Xi_c^{*}$ states will be helpful to determine whether they are conventional charm baryon or molecular state.
This idea is original from coupling of photon to molecule constituents is essentially different from that of the quark
models, in which the photon couples directly to the quark system~\cite{Koniuk:1979vy}.

Very recently,  radiative decay widths of these three states are studied by assuming them as conventional charm baryons~\cite{Wang:2020gkn,Bijker:2020tns}.
It is helpful if we could estimate the radiative decay widths to make a comparison by considering newly observed $\Xi_c^0$ states as bound state.
Thus, we can judge different explanations for structure of $\Xi_c^{*}$ when there exist experimental signals.   However, only $\Xi_c(2923)^0$
can be interpreted as a bound state~\cite{Zhu:2020jke,Jimenez-Tejero:2009cyn,Yu:2018yxl,Nieves:2019jhp} and, to date, no study addressed
 the radiation decays of $\Xi_c(2923)^0$.  In the present study, we continue our investigation of $\Xi_c(2923)^0$ properties in
terms of its radiative decay in hadronic molecule approach developed in our previous study~\cite{Zhu:2020jke}.

This paper is organized as follows. In Sec.~\ref{Sec: formulism}, we will
present the theoretical formalism. In Sec.~\ref{Sec: results}, the numerical
result will be given, followed
by discussions and conclusions in last section.

\section{THEORETICAL FORMALISM}\label{Sec: formulism}
\begin{figure}[h!]
\begin{center}
\includegraphics[bb=70 90 750 750, clip, scale=0.35]{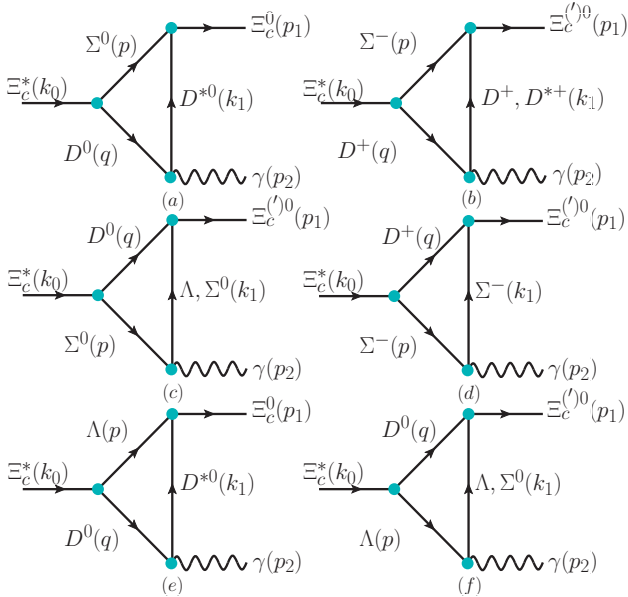}
\caption{Feynman diagrams for $\Xi_c(2930)^0\to\Xi_c^{(')0}\gamma$ decay processes.  The contributions from
$t$-channel $D^{*}$ and $D$ mesons and $\Lambda$ baryon are considered.  We also show the definition of the
kinematical ($k_0$, $p$, $q$, $p_1$, $p_2$) that we use in the present calculation.}
\label{cc1}
\end{center}
\end{figure}

In our previous study~\cite{Zhu:2020jke}, $\Xi_c(2923)^0$ was interpreted as an $S$-wave  $D\Lambda - D\Sigma$ molecular
state, in which theoretical total decay width is consistent with experimental data~\cite{LHCb:2020iby}.  In this work,
the radiative decay $\Xi_c(2923)^0\to\Xi_c^0\gamma$ are studied by assuming $\Xi_c(2923)^0$ as an $S$-wave $D\Lambda - D\Sigma$
molecular state.   The Feynman diagrams corresponding to the radiative decay are shown in Fig.~\ref{cc1}.

To compute the decay widths shown in Fig.~\ref{cc1}, we need the effective Lagrangian densities for relevant interaction
vertices.  The simplest Lagrangian densities corresponding the vertex $\Xi_c(2923)^0DY$ can be expressed as \cite{Dong:2010gu,Zhu:2020jke}
\begin{align}
{\cal{L}}_{\Xi_{c}(2923)^{0}}(x)&= \int d^4 y \Phi (y^2)   g_{\Xi_{c}(2923)^{0} D Y } D(x+\omega_Y y) \nonumber\\
	                              &\times Y (x-\omega_{D}y )\bar{\Xi}_{c}(2923)^{0}(x),\label{eq1}
\end{align}
where $\omega_{ij}=m_i/(m_i+m_j)$ with $m_{i}$ is the masses of $D$ meson or $Y$ baryon.  Since $\Sigma$ is an isovector baryon,
$Y$ should be replaced with $\vec{\Sigma} \cdot \vec{\tau}$, where $\vec{\tau}$ is the isospin matrix.  The correlation
function $\Phi (y^2)$ is introduced to describe the distribution of $D$ and $Y$ in the hadronic molecular $\Xi_c(2923)^0$ state.
The more important role of $\Phi (y^2)$ is to stop the Feynmann diagrams ultraviolet infinite.  Generally, $y$ varies from 0 to $+\infty$
and the amplitudes for the Feynman diagrams shown in Fig.~\ref{cc1} decrease to zero when $y\to\infty$.  Thus, $\Phi (y^2)$ is often
chosen to be of the following form
\begin{align}
\Phi (p^2) \doteq Exp(-p_E^2/ \Lambda ^2)
\end{align}
where $p_E$ being the Euclidean Jacobi momentum.  $\Lambda$ being the size parameter which characterizes the distribution of the components
inside the molecule and it can only be determined from experimental data.  In the following, it will be taken as a parameter and discussed later.

\begin{figure*}
\begin{center}
\includegraphics[bb=50 610 650 710, clip, scale=0.65]{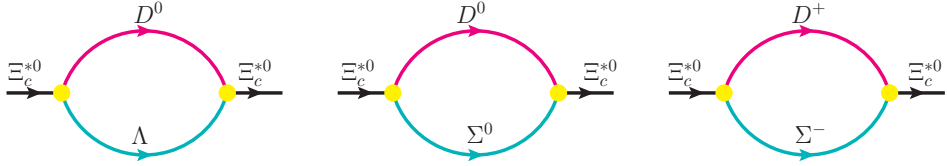}
\caption{Self energy of $\Xi_c(2923)^0$}
\label{cc}
\end{center}
\end{figure*}

The coupling constant $g_{\Xi_{c}(2923)^{0} D Y}$ is determined by the Weinberg compositeness rule~\cite{Salam:1962ap,Weinberg:1962hj},
which implies that the renormalization constant of wave function $\Xi_c(2923)^0$ is zero
 \begin{align}
 	Z_{\Xi_c(2923)^{0}} = \chi_{D \Sigma} +\chi_{D \Lambda} - \frac{d \Sigma_{\Xi_c^*}}{d k_0} \bigg| _{k\!\!\!/_0= m_{\Xi_c(2923)^{0}}} = 0,
 	\label{cou}
 \end{align}
where $\chi_{DY}$ is possibility to find $\Xi_c(2923)^0$ in the molecule state $DY$ with normalization $\chi_{D \Sigma} +\chi_{D \Lambda}=1.0$.
$\Sigma_{\Xi_c(2923)^0}$ is self-energe of $\Xi_c(2923)^0$, which can be computed through the Feynman diagrams shown in Fig.~\ref{cc}
\begin{align}
\Sigma_{\Xi_c(2923)^{0}}&(k_0)=\sum_{Y=\Lambda,\Sigma^0,\Sigma^-}  C_Y^2 g_{\Xi_c(2923)^0 D Y}^2 \int \frac{d^4 k_1}{(2\pi)^4}\nonumber\\
&\times \Phi^2[(k_1-k_0 \omega_Y)^2]  \frac{k\!\!\!/ _1 + m_Y }{k_1^2- m_Y^2}  \frac{1}{(k_1-k_0)^2-m_D^2},
\end{align}
where $k_1$ and $m_D$ is the four-momentum of baryon $Y$ and the mass of meson $D$, respectively. $k_0^2=m_{\Xi_c(2923)^0}^2$ is the
four-momentum and the mass of $\Xi_c(2923)^0$, respectively.  We set $m_{\Xi_c(2923)^0}=m_D +m_Y -E_b $ with $E_b$ being the bind energe of $\Xi_c(2923)^0$ . $C_Y$ is determined by the isospin symmetry.
\begin{align}
C_Y=\left\{
\begin{array}{cc}
	1  ~~~~&  Y=\Lambda \\
	\sqrt{2/3} ~~~~& Y=\Sigma^0\\
	-\sqrt{1/3} ~~~~& Y=\Sigma^-\\
\end{array}                   \right.
\end{align}

To compute radiative decay widthes shown in Fig.~\ref{cc}, the effective Lagrangians relevant to vertices involving a photon field are naturally
required ~\cite{Chen:2010re,Wang:2015zcp}.
\begin{align}
\mathcal{L}_{D^{*-} D^+ \gamma}&=\frac{e}{4} g_{D^{*-} D^{+} \gamma} \varepsilon^{\mu \nu \alpha \beta} F_{\mu \nu} D_{\alpha \beta}^{*-} D^{+} +H.c.,\\
\mathcal{L}_{D^{*0} D^0 \gamma}&=\frac{e}{4} g_{D^{*0} D^{0} \gamma} \varepsilon^{\mu \nu \alpha \beta} F_{\mu \nu} D_{\alpha \beta}^{*0} \bar{D}^{0} +H.c.,\\
\mathcal{L}_{D^+ D^- \gamma }&= ie A_{\mu} D^- \overleftrightarrow{\partial}^{\mu} D^+ +H.c., \\ 
\mathcal{L}_{ \gamma \Lambda \Lambda }&= \frac{e k_{\Lambda}}{2m_{\Lambda}} \bar{\Lambda} \sigma_{\mu\nu} \partial^{\nu} A^{\mu} \Lambda+H.c., 
\end{align}
\begin{align}
	\mathcal{L}_{ \gamma \Sigma \Lambda }&= \frac{e \mu_{\Sigma\Lambda}}{2m_{\Lambda}} \bar{\Sigma}^0 \sigma_{\mu\nu} \partial^{\nu} A^{\mu} \Lambda+H.c.\\
	 {\cal{L}}_{\gamma\Sigma\Sigma}&=-\bar{\Sigma}[e_{\Sigma}A\!\!\!/-\frac{e{\kappa_{\Sigma}}}{2m_{\Sigma}}\sigma_{\mu\nu}\partial^{\nu}A^{\mu}]\Sigma,	
\end{align}
where the strength tensor is defined as $F_{\mu\nu}=\partial_{\mu}A_{\nu}-\partial_{\nu}A_{\mu}$, $D^{*}_{\alpha\beta}=\partial_{\alpha}D^{*}_{\beta}-\partial_{\beta}D^{*}_{\alpha}$, and $\sigma_{\mu\nu}=i/2(\gamma_{\mu}\gamma_{\nu}-\gamma_{\nu}\gamma_{\mu})$.
The $\alpha=e^2/4\pi=1/137$ is the electromagnetic fine structure constant. The coupling constants $g_{D^{*-} D^{+} \gamma}=-0.5 Gev^{-1} $ ,$g_{D^{*0} D^{0} \gamma }=2 Gev^{-1}$~\cite{Chen:2010re,Wang:2015zcp}.  The anomalous and transition magnetic moments of the baryons are provided by the PDG~\cite{Zyla:2020zbs}
and are shown in Table~\ref{table1}.
\begin{table}[h!]
\centering
\caption{Anomalous and transition magnetic moments.
}\label{table1}
\begin{tabular}{cccccccc}
\hline\hline
~~~~~~~$\kappa_{\Sigma^{-}}=-0.16$  ~~~~~~~&$\kappa_{\Sigma^{0}}=0.65$         ~~~~~~~& $\kappa_{\Sigma^{+}}=1.46$        ~~~~~~~     \\
~~~~~~~$\kappa_{\Lambda}=-0.61$     ~~~~~~~&$u_{\Sigma\Lambda}=1.61$           ~~~~~~~&                                   ~~~~~~~    \\ \hline
\hline
\end{tabular}
\end{table}

Besides the Lagrangian above, we also require the effective Lagrangian describing coupling of vector meson to charmed baryon.
Considering the $SU(4)$ symmetry and the hidden gauge formalism, the Baryon-Baryon-Vector vertices Lagrangian is expressed as~\cite{Hofmann:2005sw,Vidana:2019amb,Zhu:2020jke}
\begin{align}
	{\mathcal{L}}_{VBB}=\frac{g_1}{4} \sum_{i,j,k,l=1}^{4} \bar{B}_{ijk} \gamma^{\mu}(V_{\mu,l}^K B^{ijl} +2 V_{\mu,l}^j B^{ilk}),
\end{align}
where coupling constant $g_1=6.6$.  $V_{\mu}$ represents 16-plet vector fileds of $\rho$, and can be expressed as
\begin{align}
	V_{\mu}&=\left(
	\begin{array}{cccc}
		\frac{1}{\sqrt{2}}(\rho^0 + \omega ) &   \rho^{+}   &  K^{*+}  & \bar{D}^{*0} \\
		\rho^{-} & \frac{1}{\sqrt{2}}(-\rho^0 + \omega ) & K^{*0} & -D^{*-}\\
		K^{*-} & \bar{K}^{*0} & \phi & D_{s}^{*-}\\
		D^{*0} & -D^{*+} & D_{s}^{*+} & J/\Psi
	\end{array}
	\right)_{\mu}.
\end{align}
 and $B$ is the tensor of baryons of the 20-plet of $p$
 \begin{align}
 	&B^{121}=p,~~B^{122}=n,~~B^{132}=\frac{1}{\sqrt{2}} \Sigma^0 - \frac{1}{6} \Lambda,\nonumber\\
 	&B^{213}=\sqrt{\frac{2}{3}} \Lambda , ~~B^{231}=\frac{1}{\sqrt{2}} \Sigma^0 +\frac{1}{\sqrt{6}} \Lambda ,~~B^{232}=\Sigma^- ,\nonumber\\
 	&B^{233}=\Xi^- , ~~B^{311}= \Sigma^+, ~~B^{313}=\Xi^0,~~B^{141}=-\Sigma_c^{++},\nonumber\\
    &B^{142}= \frac{1}{\sqrt{2}} \Sigma_c^+ +\frac{1}{\sqrt{6}} \Lambda_c,~~B^{143}=\frac{1}{\sqrt{2}} \Xi_c'^+ -\frac{1}{\sqrt{6}} \Xi_c^+,\nonumber\\
    &B^{241}=\frac{1}{\sqrt{2}} \Sigma_c^+ -\frac{1}{\sqrt{6}} \Lambda_c  ,~~B^{242}=\Sigma_c^0 ,\nonumber
 \end{align}
 \begin{align}
 	&B^{243}=\frac{1}{\sqrt{2}} \Xi_c'^0 + \frac{1}{\sqrt{6}}\Xi_c^0,~~ B^{341}=\frac{1}{\sqrt{2}} \Xi_c'^+ +\frac{1}{\sqrt{6}} \Xi_c^+ ,\nonumber\\
    &B^{124}= \sqrt{\frac{2}{3}} \Lambda_c ,~~ B^{234} =\sqrt{\frac{2}{3}} \Xi_c^0 ,~~B^{314}=\sqrt{\frac{2}{3}} \Xi_c^+ ,\nonumber\\
 	&B^{342}=\frac{1}{\sqrt{2}}\Xi_c'^0 -\frac{1}{\sqrt{6}} \Xi_c^0 , ~~B^{343}=\Omega_c^0,\nonumber\\
 	&B^{144}=\Xi_{cc}^{++}, ~~B^{244}=-\Xi_{cc}^+,~~B^{344}= \Omega_{cc},
  \end{align}
where indices $i,j,k$ of $B^{ijk}$ denote quark content of a baryon with assignments $1 \leftrightarrow \mu $,$2 \leftrightarrow d $,$3
\leftrightarrow s $,$4 \leftrightarrow c $, and first two of them is antisymmetric.

Moreover, $\Xi_{c}^{0} \Lambda D^{0}$ and $\Xi_{c}^{0} \Sigma^- D^{+}$ vertices are also required and can be obtained from Ref.~\cite{Huang:2018wgr}
\begin{align}
	{\cal{L}}_{ \Xi_{c}^{(')} \Lambda D^{}}&=\frac{i g_{\Xi_{c}^{(')} \Lambda  D}}{m_{\Xi_{c}^{(')} }+ m_{D}}  \bar{\Xi}_{c}^{(')} \gamma^{\mu} \gamma^5 \Lambda \partial^{\mu} \bar{D} +H.c.\\ 
	{\cal{L}}_{ \Xi_{c}^{(')} \Sigma D}&=\frac{i g_{\Xi_{c}^{(')} \Sigma  D}}{m_{\Xi_{c}^{(')} }+ m_{D}}  \bar{\Xi}_{c}^{(')} \gamma^{\mu} \gamma^5 \vec{\tau} \cdot \vec{\Sigma} \partial_{\mu} \bar{D} +H.c. 
\end{align}
where  $\vec{\tau}$ represents Pauli matrix,
$\vec{\Sigma}$ is $\Sigma$ triplets, and $\bar{D}$ is doublets of charmed mesons.
The relevant coupling constants are listed in Table.~\ref{table2}
\begin{table}[h!]
	\centering
	\caption{Values of relevant meson-baryon coupling constants .}
	\label{table2}
	\begin{tabular}{cccccc}
		\hline\hline
		~~~~~~~$g_{\Xi_{c} \Lambda  D}$ ~~~~~~~& $g_{\Xi_{c}'  \Lambda  D}$  ~~~~~~~&$g_{\Xi_{c} \Sigma  D}  $             ~~~~~~~& $ g_{\Xi_{c}' \Sigma  D} $  ~~~~~~~~   \\ \hline
		~~~~~~~$-5.38$    ~~~~~~~&$6.43 $         ~~~~~~~&  $ 9.31  $     ~~~~~~~& $ 3.71   $                         ~~~~~~~    \\ \hline
		\hline
	\end{tabular}
\end{table}

Putting all the piecies above together,  we obtain amplitudes for Feynman diagrams shown in Fig.~\ref{cc}.
\begin{align}
	{\cal{M}}_{a}&=\bar{\mu}(p_1)\bigg(i \frac{\sqrt{3}e}{16} C_{\Sigma^0} g_{1}g_{ \Xi_{c}^* D \Sigma }           g_{D^{*0}     D^0  \gamma } \int\frac{d^{4}k_{1}}{(2\pi)^{4}}\nonumber\\
	&\times \Phi((p\omega_{ D^{0}}-q\omega_{\Sigma^0})^2) \epsilon_{\mu\nu\alpha\beta}\gamma^{\lambda} \frac{p\!\!\!/+m_{\Sigma^{0}}}{p^2-m_{\Sigma^{0}}^2} \frac{1}{q^2-m_{D^{0}}^2} \nonumber\\
	&\times  (p_{2}^{\mu}g^{\theta\nu}-p_2^{\nu}g^{\theta\mu}) (k_1^{\alpha}g^{\beta\rho}-k_1^{\beta}g^{\alpha\rho})\frac{-g_{\lambda \rho } +k_{1 \lambda} k_{1 \rho } /m_{D^{*0}}^2}{k_{1}^{2}-m_{D^{*0}}^2}\bigg)\nonumber\\
	&\times \varepsilon^{\dagger}_{\theta}(p_2)\mu(k_0),\\      	
{\cal{M}}_{b}&=\bar{\mu}(p_1)\bigg(- i \frac{e}{(m_{\Xi_{c}^0}+m_{D^+})} C_{\Sigma^-} g_{ \Xi_{c}^* D \Sigma  } g_{\Xi_{c}^0 \Sigma^-  D^+  } \int\frac{d^{4}k_{1}}{(2\pi)^{4}}\nonumber\\
&\times\Phi((p\omega_{ D^{+}}-q\omega_{\Sigma^-})^2) k\!\!\!/_1 	\gamma^5 \frac{p\!\!\!/+m_{\Sigma^{-}}}{p^2-m_{\Sigma^{-}}^2}
	\frac{1}{q^2-m_{D^{+}}^2} (q^{\theta}+k_1^{\theta}) \nonumber
\end{align}
\begin{align}
&\times  \frac{1}{k_{1}^{2}-m_{D^{+}}^2} -i \frac{\sqrt{6}e}{16}  C_{\Sigma^-} g_{1}g_{ \Xi_{c}^* D \Sigma } g_{D^{*+} D^+  \gamma } \int\frac{d^{4}k_{1}}{(2\pi)^{4}}  \nonumber\\
&\times \Phi((p\omega_{D^{+}}-q\omega_{\Sigma^-})^2)\epsilon_{\mu\nu\alpha\beta}\gamma^{\lambda} \frac{p\!\!\!/+m_{\Sigma^{-}}}{p^2-m_{\Sigma^{-}}^2} \frac{1}{q^2-m_{D^{+}}^2}\nonumber\\
&\times (p_{2}^{\mu}g^{\theta\nu}-p_2^{\nu}g^{\theta\mu}) (k_1^{\alpha}g^{\beta\rho}-k_1^{\beta}g^{\alpha\rho})\frac{-g_{\lambda \rho } +k_{1 \lambda} k_{1 \rho } /m_{D^{*+}}^2}{k_{1}^{2}-m_{D^{*0}}^2}\bigg)\nonumber\\
	&\times
	\varepsilon^{\dagger}_{\theta}(p_2)\mu(k_0), \\	
	{\cal{M}}_{b}'&=\bar{\mu}(p_1)\bigg(- i \frac{e}{(m_{\Xi_{c}'^0}+m_{D^+})} C_{\Sigma^-} g_{ \Xi_{c}^* D \Sigma  }    g_{\Xi_{c}'^0 \Sigma^-  D^+  } \int\frac{d^{4}k_{1}}{(2\pi)^{4}}\nonumber\\
	&\times\Phi((p\omega_{ D^{+}}-q\omega_{\Sigma^-})^2) k\!\!\!/_1 	\gamma^5 \frac{p\!\!\!/+m_{\Sigma^{-}}}{p^2-m_{\Sigma^{-}}^2}
	\frac{1}{q^2-m_{D^{+}}^2} (q^{\theta}+k_1^{\theta}) \nonumber\\
	&\times  \frac{1}{k_{1}^{2}-m_{D^{+}}^2} \bigg)
	\varepsilon^{\dagger}_{\theta}(p_2)\mu(k_0), \\	
{\cal{M}}_{c}^{(')}&=\bar{\mu}(p_1)\bigg(i  \frac{e \mu_{\Sigma \Lambda}}{4 m_{\Lambda}(m_{\Xi_{c}^{(')0}}+m_{D^0})} C_{\Sigma^0} g_{ \Xi_{c}^* D \Sigma } g_{\Xi_{c}^{(')0} \Lambda D^0} \int\frac{d^{4}k_{1}}{(2\pi)^{4}}\nonumber\\
&\times\Phi((q\omega_{ \Sigma^{0}}-p \omega_{D^0})^2) q\!\!\!/ \gamma^5  \frac{k\!\!\!/_1 +m_{\Lambda}}{k_{1}^{2}-m_{\Lambda}^2} (\gamma^{\theta} p\!\!\!/ _2 - p\!\!\!/ _2  \gamma^{\theta})    \nonumber\\
&\times   \frac{p\!\!\!/ +m_{\Sigma^0}}{p^2-m_{\Sigma^{0}}^2}  \frac{1}{q^2-m_{D^{0}}^2} +i  \frac{e k_{\Sigma}}{4 m_{\Sigma}(m_{\Xi_{c}^{(')0}}+m_{D^0})} \nonumber\\
& \times C_{\Sigma^0} g_{ \Xi_{c}^* D \Sigma } g_{\Xi_{c}^{(')0} \Sigma^0 D^0} \int\frac{d^{4}k_{1}}{(2\pi)^{4}} \Phi((q\omega_{ \Sigma^{0}}-p \omega_{D^0})^2) \nonumber\\
	&\times q\!\!\!/ \gamma^5  \frac{k\!\!\!/_1 +m_{\Sigma^0}}{k_{1}^{2}-m_{\Sigma^0}^2} (\gamma^{\theta} p\!\!\!/ _2 - p\!\!\!/ _2  \gamma^{\theta})   \frac{p\!\!\!/ +m_{\Sigma^0}}{p^2-m_{\Sigma^{0}}^2}  \frac{1}{q^2-m_{D^{0}}^2}\bigg)  \nonumber\\
	&\times
	\varepsilon^{\dagger}_{\theta}(p_2)\mu(k_0),\\  
{\cal{M}}_{d}^{(')}&=\bar{\mu}(p_1)\bigg(-i  \frac{e }{m_{\Xi_{c}^{(')0}}+m_{D^+}}  C_{\Sigma^-} g_{ \Xi_{c}^{*0} D^+ \Sigma^- } g_{\Xi_{c}^{(')0} D^+ \Sigma^- } \int\frac{d^{4}k_{1}}{(2\pi)^{4}}    \nonumber\\
	&\times  \Phi((q\omega_{ \Sigma^{-}}-p\omega_{D^+})^2) q\!\!\!/ \gamma^5  \frac{k\!\!\!/_1 +m_{\Sigma^-}}{k_{1}^{2}-m_{\Sigma^-}^2} (\gamma^{\theta} + \frac{k_{\Sigma^-}}{4 m_{\Sigma}}(\gamma^{\theta} p\!\!\!/ _2   \nonumber\\
	&- p\!\!\!/ _2  \gamma^{\theta}))
	\frac{p\!\!\!/ +m_{\Sigma^-}}{p^2-m_{\Sigma^{-}}^2}  \frac{1}{q^2-m_{D^{+}}^2}\bigg)
	\varepsilon^{\dagger}_{\theta}(p_2)\mu(k_0),\\ 
{\cal{M}}_{e}&=\bar{\mu}(p_1)\bigg( i \frac{e}{16}  C_{\Lambda } g_{1}g_{ \Xi_{c}^* D \Lambda } g_{D^{*0} D^0  \gamma } \int\frac{d^{4}k_{1}}{(2\pi)^{4}}\nonumber\\
	&\times  \Phi((p\omega_{ D^{0}}-q\omega_{\Lambda})^2) \epsilon_{\mu\nu\alpha\beta}\gamma^{\lambda} \frac{p\!\!\!/+m_{\Lambda}}{p^2-m_{\Lambda}^2} \frac{1}{q^2-m_{D^{0}}^2} \nonumber\\
	&\times (p_{2}^{\mu}g^{\theta\nu}-p_2^{\nu}g^{\theta\mu}) (k_1^{\alpha}g^{\beta\rho}-k_1^{\beta}g^{\alpha\rho})\frac{-g_{\lambda \rho } +k_{1 \lambda} k_{1 \rho } /m_{D^{*0}}^2}{k_{1}^{2}-m_{D^{*0}}^2}\bigg)\nonumber\\
	& \times 	\varepsilon^{\dagger}_{\theta}(p_2)\mu(k_0),\\ 
{\cal{M}}_{f}^{(')}&=\bar{\mu}(p_1)\bigg(i  \frac{e k_{\Lambda}}{4 m_{\Lambda}(m_{\Xi_{c}^{(')0}}+m_{D^0})}  C_{\Lambda } g_{ \Xi_{c}^* D \Lambda } g_{\Xi_{c}^{(')0} \Lambda D^0} \int\frac{d^{4}k_{1}}{(2\pi)^{4}}\nonumber\\
	&\times \Phi((q\omega_{ \Lambda}-p\omega_{D^0})^2) q\!\!\!/ \gamma^5 \frac{k\!\!\!/_1 +m_{\Lambda}}{k_{1}^{2}-m_{\Lambda}^2} (\gamma^{\theta} p\!\!\!/ _2 - p\!\!\!/ _2  \gamma^{\theta})   \nonumber\\
	&\times   \frac{p\!\!\!/ +m_{\Lambda}}{p^2-m_{\Lambda}^2}  \frac{1}{q^2-m_{D^{0}}^2}  +i  \frac{e \mu_{\Sigma\Lambda}}{4 m_{\Lambda}(m_{\Xi_{c}^{(')0}}+m_{D^0})}  \nonumber\\
	& \times  C_{\Lambda } g_{ \Xi_{c}^* D \Lambda } g_{\Xi_{c}^{(')0} \Sigma^0 D^0} \int\frac{d^{4}k_{1}}{(2\pi)^{4}}     \Phi((q\omega_{ \Lambda}-p\omega_{D^0})^2)  \nonumber
\end{align}
\begin{align}
	&\times  q\!\!\!/ \gamma^5 \frac{k\!\!\!/_1 +m_{\Sigma^0}}{k_{1}^{2}-m_{\Sigma^0}^2}(\gamma^{\theta} p\!\!\!/ _2 - p\!\!\!/ _2  \gamma^{\theta})  \frac{p\!\!\!/ +m_{\Lambda}}{p^2-m_{\Lambda}^2}  \nonumber\\
	&\times    \frac{1}{q^2-m_{D^{0}}^2}\bigg)
	\varepsilon^{\dagger}_{\theta}(p_2)\mu(k_0)
\end{align}
where ${\cal{M}}$ and ${\cal{M}}^{'}$ is the amplitude for $\Xi_c(2923)^0 \rightarrow \Xi_c^{0} \gamma$ and $ \Xi_c(2923)^0 \rightarrow \Xi_c^{'0} \gamma$, respectively.

\begin{figure}[h!]
\begin{center}
\includegraphics[bb=120 610 500 710, clip, scale=0.70]{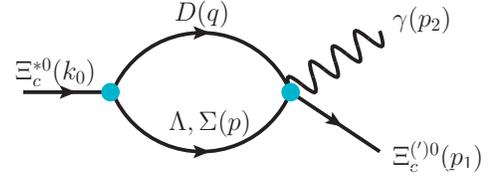}
\caption{Contact diagram for $\Xi_c(2923)^0 \rightarrow \Xi_c^{(')0} \gamma$. We also indicate definitions of the kinematics ($p_1,p_2,k_1,k_2$,and$ p$) used in the calculation}
\label{cc2}
\end{center}
\end{figure}

Summing up all the individual amplitudes ,we obtain the total amplitude of $ \Xi_c(2923)^0 \rightarrow \Xi_c^{(')0} \gamma$ .
\begin{align}
	\mathcal{M}^T(\Xi_c(2923)^0 \rightarrow \Xi_c^{(')0} \gamma)&=\mathcal{M}_a+\mathcal{M}_b^{(')}+\mathcal{M}_c^{(')}+\mathcal{M}_d^{(')}\nonumber\\
	&+\mathcal{M}_e+\mathcal{M}_f^{(')}
\end{align}
However , the $\mathcal{M}^T$ we obtained currently is not satisfying the gauge invariance of the photon filed.  Therefore,
the contact diagram in Fig.~\ref{cc2} must be included to ensure the gauge invariance of total amplitudes.  we employ the following
form to satisfy $p_2^{\theta} \mathcal{M}_{\theta}^{total}(\equiv \mathcal{M}^{T}+ \mathcal{M}_{com})=0$.
\begin{align}
	&{\cal{M}}_{com}(\Xi_c^{*0}\to\gamma\Xi_c^{(')0})=i  \frac{e}{m_{\Xi_c^{(')0}} +m_{D^+}}C_{\Sigma^-}g_{\Xi_c^{*0}\Sigma^{-}D^{+}}g_{\Xi_c^{(')0}\Sigma^{-}D^{+}} \nonumber\\
	&\times \int_0^{\infty}d\alpha \int_0^{\infty}d\eta \int_0^{\infty}d\zeta\frac{1}{16\pi^2y^2\beta^2}\bar{u}(p_1)({\cal{C}}_1^{\theta}{\cal{T}}_1+{\cal{C}}_2^{\theta}{\cal{T}}_2)\gamma_5  \nonumber\\
	&\times u(k_0)\epsilon^{*}_{\theta}(p_2),
\end{align}
Where
\begin{align}
	{\cal{T}}_1&=\exp\{-\frac{1}{\beta^2}[\eta(-p_2^2+m^2_{D^{+}})+\alpha{}m_{D^{+}}^2+\zeta(-p_1^2+m^2_{\Sigma^{-}})                    \nonumber\\
                	&-(p_1\omega_{D^{+}}-p_2\omega_{\Sigma^{-}})^2+\frac{1}{4y}({\cal{H}}_2p_2-{\cal{H}}_1p_1)^2]\},\\ 
{\cal{T}}_2&=\exp\{-\frac{1}{\beta^2}[\eta(-p_1^2+m^2_{D^{+}})+\alpha{}m_{\Sigma^{-}}^2+\zeta(-p_2^2+m^2_{\Sigma^{-}})                  \nonumber\\
	         &-(p_1\omega_{\Sigma^{-}}-p_2\omega_{D^{+}})^2+\frac{1}{4y}({\cal{H}}_2p_1-{\cal{H}}_1p_2)^2]\},\\ 
{\cal{C}}_1^{\theta}&=-\frac{m_1^2{\cal{H}}_1^3}{4y^3}p_1^{\theta}   -\frac{{\cal{H}}_1^2 {\cal{H}}_2}{4 y^3}(mm_1-m_1^2) p_1^{\theta} -\frac{{\cal{H}}_1^2 {\cal{H}}_2}{4 y^3 } \nonumber\\
&\times (m^2 -mm_1) p_1^{\theta} +\frac{{\cal{H}}_1^2 m_{\Sigma^- } m_1 }{2 y^2 } p_1 ^{\theta } -\frac{{\cal{H}}_1^2 m_1^2 }{2 y^2} p_1^{\theta} \nonumber\\
	& - \frac{{\cal{H}}_1 {\cal{H}}_2 m_{\Sigma^- }}{ 2 y^2 } (m - m_1 )p_1^{\theta } + \frac{{\cal{H}}_1 {\cal{H}}_2 }{2 y^2 } (m^2- m m_1 ) p_1^{\theta }\nonumber\\	
	& - \frac{{\cal{H}}_1 }{2 y^2 }(2 p_1^{\theta} - m_1 \gamma^{\theta }) - \frac{{\cal{H}}_1 m_1 }{2 y^2 } \gamma^{\theta } - \frac{2 {\cal{H}}_1 }{y^2} p_1^{\theta} - \frac{m_{\Sigma^-}}{y} \gamma^{\theta} \nonumber
 \end{align}
        \begin{align}
	& + \frac{1}{y}(2 p_1^{\theta} -m_1 \gamma^{\theta}),\\ 
{\cal{C}}_2^{\theta}&= \frac{{\cal{H}}_1 {\cal{H}}_2 m_{\Sigma^- } }{ 4 y^2 }( m^2 - m m_1 )\gamma^{\theta}    -\frac{{\cal{H}}_1 {\cal{H}}_2 }{2 y^2 }(m^2 -mm_1 ) p_1^{\theta }  \nonumber\\
	&+ \frac{{\cal{H}}_1 {\cal{H}}_2 }{2 y^2 } p_1 \cdot p_2 ( 2 p_1 ^{\theta} - m_1 \gamma^{\theta } ) -\frac{{\cal{H}}_2 ^2 {\cal{H}}_1 }{4 y^3 } p_1 \cdot p_2 (2 p_1 ^{\theta}  \nonumber\\
	& - m_1 \gamma^{\theta} ) +\frac{m_{\Sigma^- }}{y } \gamma^{\theta} - \frac{3 {\cal{H}}_2}{2 y^2 }(2 p_1 ^{\theta} - m_1 \gamma^{\theta}) - m_{\Sigma^-}^2 m_1 \gamma^{\theta}  \nonumber\\
	&+\frac{{\cal{H}}_2 m_{\Sigma^- }^2 m_1}{2 y} \gamma^{\theta} -\frac{m_1 }{y} \gamma^{\theta} +\frac{{\cal{H}}_2 m_{\Sigma^- } m_1 }{y } p_1^{\theta} - \frac{{\cal{H}}_2 ^2 m_{\Sigma^-} m_1}{2 y^2 } p_1^{\theta}  \nonumber\\
	& - \frac{{\cal{H}}_2 ^2 m_1^2 }{4 y^2 }(2 p_1^{\theta} -m_1 \gamma^{\theta} ) +\frac{{\cal{H}}_2^3 m_1^2 }{8 y^3}(2 p_1^{\theta} -m_1 \gamma^{\theta})  , 
\end{align}
where $y=1+\alpha+\eta+\zeta$,${\cal{H}}_2=2(\eta+\omega_{\Sigma^{-}})$, ${\cal{H}}_1=2(\zeta+\omega_{D^{+}})$.
$m$ and $m_1$ are the masses of $\Xi_c^{*0}$ and $\Xi_c^{(')0}$, respectively.

Once the amplitudes are determined, the corresponding partial width can be obtained immediately with the following formula
\begin{align}
	d\Gamma(\Xi_c(2923)^0 \rightarrow \Xi_c^{(')0} \gamma)=\frac{1}{2J+1} \frac{1}{32\pi^2} \frac{|\vec{p}_1|}{m_{\Xi_c^{*0}}^2} |\bar{\mathcal{M}}|^2 d \Omega.
\end{align}
Where $J$ is the total angular momentum of $\Xi_c(2923)^0$, $|\vec{p}_1|$ is the three-momentum of the decay products in the center of mass frame,
and the overline indicates the sum over the polarization vectors of the final hadrons.

\section{RESULTS}\label{Sec: results}
To make a reliable prediction for the radiative decay widths of $\Xi_c(2923)^0$, two issues we need to clarify are, respectively,
the relation of the parameter $\Lambda$ to the correlation function $\Phi (y^2)$ and the coupling of $\Xi_c(2923)^0$ with its molecular
compositions.  Unfortunately, the parameter $\Lambda$ could not be determined by first principles, it is usually set to be about 1.0 GeV
to reproduce the experimentally observed decay width in the literature~\cite{Zhu:2020jke,Dong:2010gu,Yang:2021pio,Zhu:2021exs,Huang:2018wgr,Dong:2009uf}
(and their references).  In this work, we adopt the value $\Lambda=1.0$ GeV because it is determined from the experimental data of
Refs.~~\cite{Zhu:2020jke,Dong:2010gu,Yang:2021pio,Zhu:2021exs,Huang:2018wgr,Dong:2009uf}(and their references) within the same correlation function $\Phi (y^2)$
adopted in current work.

\begin{figure}[h!]
	\centering
	\includegraphics[bb=00 00 680 515,clip,scale=0.33]{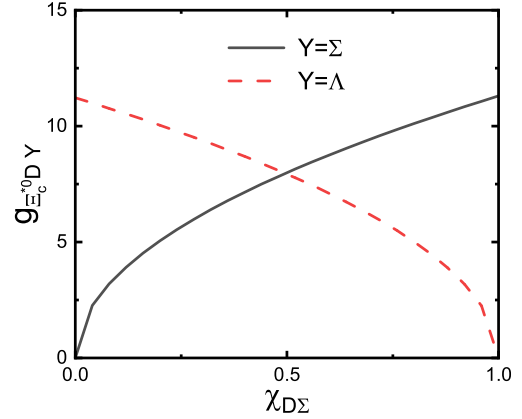}
	\caption{(color online) The coupling constant of $g_{\Xi_{c}^{*0} D Y}$ as a function of the parameter $\chi_{D\Sigma}$.}
	\label{fig.coupling}
\end{figure}
The coupling constants corresponding to the effective Lagrangians  listed in Eq.~(\ref{eq1}) have been determined.
Although detailed results can find in Ref.~\cite{Zhu:2020jke}, we will review them here again.  With a value of
$\chi_{D\Sigma}$=0.0-1.0 and the compositeness condition that we introduced in Eq,~(\ref{cou}), the $\chi_{D\Sigma}$ dependence
of the coupling constants $g_{\Xi_c^{*0}D\Lambda}$ and $g_{\Xi_c^{*0}D\Sigma}$ are computed and are shown in Fig.~\ref{fig.coupling}.
We find that the coupling constant $g_{\Xi_c^{*0}D\Lambda}$ monoto-nously decreases with increasing $\chi_{D\Sigma}$, while the
coupling constant $g_{\Xi_c^{*0}D\Sigma}$ increases with increasing $\chi_{D\Sigma}$.  The opposite trend can be easily understood,
as the coupling constants $g_{\Xi_c^{*0}D\Lambda}$ and $g_{\Xi_c^{*0}D\Sigma}$ are directly proportional to the corresponding molecular
compositions~\cite{Dong:2009uf,Huang:2020taj}.  And a simple relation between the coupling constants $g_{\Xi_c^{*0}D\Lambda}$ and
$g_{\Xi_c^{*0}D\Sigma}$ can be deduced as
\begin{align}
({\cal{A}}g_{\Xi_c^{*0}D\Sigma})^2=1-({\cal{B}}g_{\Xi_c^{*0}D\Lambda})^2
\end{align}
with ${\cal{A}}=0.0885$ and ${\cal{B}}=0.0891$.

\begin{figure}[h!]
	\centering
	\includegraphics[bb=00 20 700 530, clip, scale=0.33]{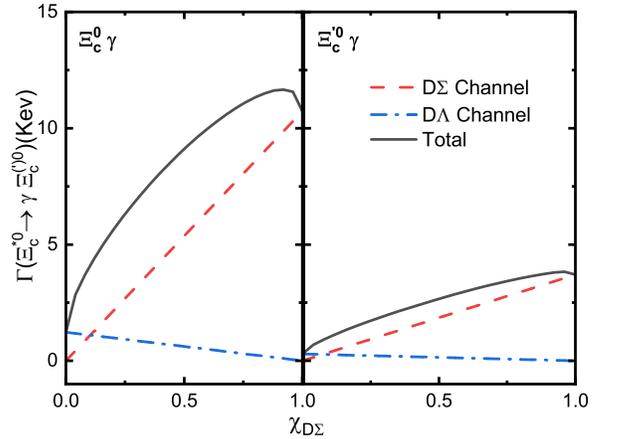}
	\caption{(color online) The partial decay widths from $D\Sigma$ and $D \Lambda$ contribution as a function of the parameter $\chi_{D\Sigma}$. }
	\label{fig.component}
\end{figure}
Once the model parameter $\Lambda$ and coupling constants $g_{\Xi_c^{*0}D\Lambda}$ and $g_{\Xi_c^{*0}D\Sigma}$ are determined,
the radiative decay width $\Xi_c(2923)^0\to\Xi_c^{(')0}\gamma$ can be calculated straightforwardly.  In Fig.~\ref{fig.component},
the radiative decay width $\Xi_c(2923)^0\to\Xi_c^{(')0}\gamma$ verses $\chi_{D\Sigma}$ is presented, where we restrict the value of $\chi_{D\Sigma}$
from 0.0 to 1.0.   The results show that the value of the radiative decay width $\Xi_c(2923)^0\to\Xi_c^{(')0}\gamma$ are the smallest for
$\chi_{D\Sigma}=0.0$ case, in this case $\Xi_c(2923)^0$ is a pure $D\Lambda$ molecular state.  With the increase of $\chi_{D\Sigma}$,
the radiative decay width $\Xi_c(2923)^0\to\Xi_c^{(')0}\gamma$ increases.  However, the decay width for $\Xi_c(2923)^0\to\Xi_c^{0}\gamma$ and
$\Xi_c(2923)^0\to\Xi_c^{'0}\gamma$ decreases when $\chi_{D\Sigma}$ varies from 0.92 to 1.0 and 0.94 to 1.0, respectively.
Such dependence of the radiative decay width on $\chi_{D\Sigma}$ can be easily understood due to the interference between
$D\Lambda$ and $D\Sigma$ channels is sizable (see Fig.~\ref{fig.component}), leading to a total radiative decay width is the largest.
It means that the $D\Lambda$ channel strongly couples to the $D\Sigma$ channel, and the same conclusion can also find in
Refs.~\cite{Zhu:2020jke, Jimenez-Tejero:2009cyn,Yu:2018yxl, Nieves:2019jhp}.

The contributions of the $D\Lambda$ and $D\Sigma$ components for $\Xi_c(2923)^0\to\Xi_c^{(')0}\gamma$ decays
are calculated and also presented in Fig.~\ref{fig.component}.  We find that the $D\Lambda$ channel contribution decreases with
increasing $\chi_{D\Sigma}$, while the contribution from $D\Sigma$ channel increases with increasing $\chi_{D\Sigma}$.  Moreover,
the $D\Lambda$ channel plays a predominant role at small $\chi_{D\Sigma}$, while the contribution  from the $D\Sigma$ channel
becomes the most important when $\chi_{D\Sigma}$ is larger than 0.11 for the process $\Xi_c(2923)^0\to\Xi_c^{0}\gamma$ and 0.07
for the process $\Xi_c(2923)^0\to\Xi_c^{'0}\gamma$.

\begin{figure}[h!]
\centering
\includegraphics[bb=00 15 800 530, clip, scale=0.33]{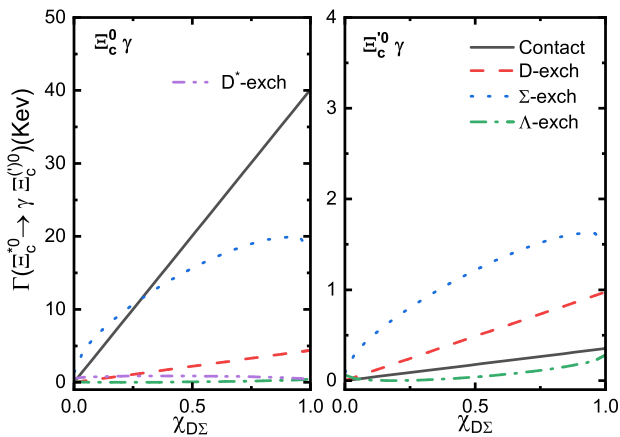}
\caption{(color online) The partial decay widths from $D$(red dash  line), $\Sigma$(blue  dot  line), $D^*$ (purple dash dot dot line),
$\Lambda$(green dash dot  line) exchange contribution and the contact term (black line) for the $\Xi_c(2923)^0\to\gamma\Xi^{(')0}$ as a
function of the parameter $\chi_{D\Sigma}$.}\label{fig.exchange}
\end{figure}
The individual contributions of the $D$, $D^{*}$, $\Sigma$, $\Lambda$ exchanges, and the contact term for $\Xi_c(2923)^{0}\to\Xi_c^{(')0}\gamma$ decays
are calculated and shown in Fig.~\ref{fig.exchange}.  According to the Lagrangian above, the relative signs of the Feynman diagrams shown
in Fig.~(\ref{cc1}) are well defined.  And the total decay widths obtained are the square of their coherent sum.   We can find that the $\Sigma$ exchange
and contact term play a dominant role, while the $D$, $D^{*}$, and $\Lambda$ exchanges give minor contributions.  However, as $\chi_{D\Sigma}$ gets bigger,
the contribution from the contact term becomes the most important.  The interferences among them are still sizable.  And it makes the total
radiative decay width at the order of 1.23-11.66 KeV.  For $\Xi_c(2923)^{0}\to\Xi_c^{'0}\gamma$ transition, the predicted total decay width increases
from 0.30 to 3.71 KeV.  The $\Sigma$ and $D$ exchanges provide the dominant contribution, while the contact term contribution that play a dominant role in the $\Xi_c(2923)^{0}\to\Xi_c^{0}\gamma$ reaction is small.  The sizable interference between them still exists.   We also find that the $D^{*}$
contribution that is considered in the $\Xi_c(2923)^{0}\to\Xi_c^{0}\gamma$ reaction is not included for the process $\Xi_c(2923)^{0}\to\Xi_c^{'0}\gamma$.

It is interesting to compare our results with those in Ref.~\cite{Wang:2020gkn,Bijker:2020tns}.  According to Ref.~\cite{Bijker:2020tns},
the $\Xi_c(2923)^0$ may be conventional charmed baryons with $P$-wave.  The electromagnetic decay width $\Xi_c(2923)^0\to{}\Xi_c^0\gamma$
is found to be small and  of the order of 1.1 KeV, whereas the decay width is about 742.0 KeV and 0.4 KeV for the transition
$\Xi_c(2923)^0\to{}^{2}\Xi^{'0}_c\gamma$ and $^{4}\Xi^{'0}_c\gamma$, respectively.  The conclusion of the conventional charmed baryon for
$\Xi_c(2923)^0$ is also supported by Ref~\cite{Wang:2020gkn}.  They find that the radiative decay width of $\Xi_c(2923)^0$ into the ground
state $^{2}\Xi_c$ is about 0.0 KeV, while the transition $\Xi_c(2923)^0\to{}^{2}\Xi^{'0}_c\gamma$ play a dominant role.  These are quite
different from our results that the total radiative decay widths for $\Xi_c(2923)^0\to{}\Xi^{0}_c\gamma$ and $\Xi_c(2923)^0\to{}\Xi^{'0}_c\gamma$
is of the order of 1.23-11.66 KeV and 0.30-3.71 KeV, respectively, by assuming $\Xi_c(2923)^0$ as an $S$-wave $D\Lambda-D\Sigma$ bound state.
Reader can find these differentia in Tab.~\ref{table3}.   If measurements are in future experimental, these differences will be very useful to help
us to test various interpretations of $\Xi_c(2923)^0$.
\begin{table}[h!]
\centering
\caption{Radiative decay widths of $\Xi_c(2923)^0$ in KeV.}\label{table3}
\begin{tabular}{cccccccc}
\hline\hline
~~~$\Xi_c(2923)^0 \to$        ~~~~~~~&$^{2}\Xi_c^{'}+\gamma$    ~~~~~~~& $^{4}\Xi_c^{'}+\gamma$        ~~~~~~~ & $^{2}\Xi_c+\gamma$~~~    \\
~~~Ref.~\cite{Wang:2020gkn}   ~~~~~~~&$472.0$                   ~~~~~~~& $1.0$                         ~~~~~~~ & $0.0$             ~~~    \\
~~~Ref.~\cite{Bijker:2020tns} ~~~~~~~&$451.0$                   ~~~~~~~& $0.4$                         ~~~~~~~ & $1.1$             ~~~    \\
~~~ This work                 ~~~~~~~&$0.30-3.71$                     ~~~~~~~& $0.0$                         ~~~~~~~ & $1.23-11.66$             ~~~    \\ \hline
\hline
\end{tabular}
\end{table}

\section{Summary}
Presently, there is not sufficient experimental information to determine the spin-parity of $\Xi_c(2923)^0$ state.
Its properties, however, such as the spectroscopy and the strong decay widths, can be well explained in the context of
the conventional charm baryon~\cite{Wang:2020gkn,Lu:2020ivo,Yang:2020zjl,Agaev:2020fut,Hu:2020zwc} state or molecular
state~\cite{Zhu:2020jke,Jimenez-Tejero:2009cyn,Yu:2018yxl,Nieves:2019jhp}.  Precise information on the radiative decay mechanism
of $\Xi_c(2923)^0$ will be helpful to determine whether it is a conventional charm baryon or molecular state.
Fortunately, the radiative decay widths of$\Xi_c(2923)^0$ have been studied by assuming $\Xi_c(2923)^0$ as conventional
charm baryon~\cite{Wang:2020gkn,Bijker:2020tns}.  It is helpful if we could estimate the radiative decay width to make a comparison
by assuming $\Xi_c(2923)^0$ as a molecular state.  Thus, we can judge the different explanations for the structure of
$\Xi_c(2923)^0$ if there exist experimental signals.

In the present study, we estimated the partial widths for the radiative decay from the $\Xi_c(2923)^0$ to the
$\Xi_c^0$ and $\Xi_c^{'0}$ state in a molecular scenario, in which the $\Xi_c(2923)^0$ is considered a $D\Lambda-D\Sigma$
hadronic molecule.  In the relevant parameter region, the partial widths are evaluated as
\begin{align}
\Gamma(\Xi_c(2923)^0\to\Xi_c^0\gamma)=1.23-11.66 ~~~{\rm KeV},\nonumber\\
\Gamma(\Xi_c(2923)^0\to\Xi_c^{'0}\gamma)=0.30-3.71 ~~~{\rm KeV}.\nonumber
\end{align}
Our estimations indicate that the partial widths for the
transition $\Xi_c(2923)^0\to\Xi_c^{'0}\gamma$ are approximately one order of magnitude smaller than those of $\Xi_c(2923)^0\to\Xi_c^{0}\gamma$.
Our results are quite different from the results~\cite{Wang:2020gkn,Bijker:2020tns} that are obtained by assuming $\Xi_c(2923)^0$ may be conventional
charmed baryon.  And can be used to test the (molecular) nature of the $\Xi_c(2923)^0$.

\section*{Acknowledgments}
This work was supported by the National Natural Science Foundation of China under Grant No.12104076,
the Science and Technology
Research Program of Chongqing Municipal Education Commission (Grant No. KJQN201800510), and the Opened Fund
of the State Key Laboratory on Integrated Optoelectronics
(GrantNo. IOSKL2017KF19).  Yin Huang want to thanks
the support from the Development and Exchange Platform for
the Theoretic Physics of Southwest Jiaotong University under Grants No.11947404 and No.12047576,
the Fundamental Research Funds for the Central Universities(Grant No.
2682020CX70), and the National Natural Science Foundation
of China under Grant No.12005177.

\end{document}